\documentclass[twocolumn,showpacs,amsmath,amssymb]{revtex4}
\usepackage{graphicx}
\usepackage{dcolumn}
\usepackage{bm}
\usepackage{slashbox}
\newcommand{\BR}{{\cal B}}

\newcommand{\piz}{\pi^0}

\newcommand{\EE}{e^+e^-}

\newcommand{\pp}{\pi^+\pi^-}

\newcommand{\beq}{\begin{equation}}
\newcommand{\eeq}{\end{equation}}
\newcommand{\beqn}{\begin{eqnarray}}
\newcommand{\eeqn}{\end{eqnarray}}
\newcommand{\beqns}{\begin{eqnarray*}}
\newcommand{\eeqns}{\end{eqnarray*}}
\newcommand{\bfg}{\begin{figure}}
\newcommand{\efg}{\end{figure}}
\newcommand{\bitm}{\begin{itemize}}
\newcommand{\eitm}{\end{itemize}}
\newcommand{\bnum}{\begin{enumerate}}
\newcommand{\enum}{\end{enumerate}}
\newcommand{\btbl}{\begin{table}}
\newcommand{\etbl}{\end{table}}
\newcommand{\btbu}{\begin{tabular}}
\newcommand{\etbu}{\end{tabular}}
\def\eref#1{(\ref{#1})}

\def\prd#1#2#3 {{~Phys. Rev. D {#1}, #2 (#3) }} 
\def\plb#1#2#3 {{~Phys. Lett. B {#1}, #2 (#3) }} 

\begin{document}
\preprint{Draft-v1}

\title{Branching fraction of the isospin violating process
$\phi\to \omega\piz$}
\author{C.~Z.~Yuan}
 \email{yuancz@ihep.ac.cn}
\author{X.~H.~Mo}
 \email{moxh@ihep.ac.cn}
\author{P.~Wang}
 \email{wangp@ihep.ac.cn}
\affiliation{Institute of High Energy Physics, Chinese Academy of
Sciences, Beijing 100049, China}

\date{\today}

\begin{abstract}

We examine the parametrization of the $\EE\to \omega\piz$ cross
section in the vicinity of the $\phi$ resonance and the extraction
of the branching fraction of the isospin violating process
$\phi\to \omega\piz$ from experimental data. We found that there
are two possible solutions of the branching fraction, one is
$4\times 10^{-5}$, and the other is $7\times 10^{-3}$. The latter
is two orders of magnitude higher than the former, which is the
commonly accepted one.

\end{abstract}


\maketitle

\section{Introduction}

As has been pointed out in a recent study~\cite{Yuan:2009gd},
there are many cases where multiple solutions are found in fitting
one dimensional distribution with the coherent sum of several
amplitudes and free relative phase between them. The fit to the
$\EE\to \omega\piz$ cross sections in the vicinity of the $\phi$
resonance was shown as an example of the existence of the two
solutions and how large the difference could be between them.

However, Ref.~\cite{Yuan:2009gd} found these two solutions only
through a fit to the experimental data, which causes suspicion
that the two solutions may due to the statistical fluctuation or
other reasons associated with the data handling or fitting
procedure or something else. In this brief report, we show
mathematically that two solutions exist in the parametrization of
the cross section used in the original
publication~\cite{kloe_omegapi}; and the second solution can be
obtained analytically from the solution reported in the literature
without doing fit to the experimental data.

\section{Derive of the second solution}

The cross section of $\EE\to \omega\piz$ as a function of the
center-of-mass energy, $\sqrt{s}$, is parameterized as
 \beq
 \label{xsection}
\sigma(\sqrt{s}) = \sigma_{nr}(\sqrt{s})
\cdot\left|1-Z\frac{M_{\phi}\Gamma_{\phi}}{D_{\phi}(\sqrt{s})}\right|^2
 \eeq
in Ref.~\cite{kloe_omegapi}, where $\sigma_{nr}(\sqrt{s}) =
\sigma_{0} + \sigma' (\sqrt{s} - M_\phi)$ is the bare cross
section for the non-resonant process, parameterized as a linear
function of $\sqrt{s}$; $M_{\phi}$, $\Gamma_{\phi}$, and
$D_{\phi}=M_{\phi}^2-s-iM_{\phi}\Gamma_{\phi}$ are the mass, the
width, and the inverse propagator of the $\phi$ meson,
respectively. Here $Z$ is a complex number which depicts the
interference effect. Conventionally, the real and imaginary parts
of $Z$ are denoted as $\Re(Z)$ and $\Im(Z)$, respectively.

If we write
 \beq
G(s,Z) =1-Z \frac{M_{\phi}\Gamma_{\phi}}{D_{\phi}(\sqrt{s})}~,
\label{defgsz}
 \eeq
then in the complex-parameter space (denoted by a complex number
$Z'$), we want to figure out all possible parameters which can
satisfy the following relation
 \beq
\left| G(s,Z) \right|^2 = \left| G(s,Z') \right|^2~.
\label{gsqrelation}
 \eeq
Notice that the above relation is to be true for any $s$, it
should be true for some special values of $s$. If we firstly take
a special value of $s$ which satisfies $M_{\phi}^2-s=0$, then we
obtain
 \beq
 |1-iZ|^2 = |1-iZ'|^2~,
 \eeq
or
 \beq
 \label{eq3}
 |Z|^2+2\Im(Z)=|Z'|^2+2\Im(Z')~.
 \eeq
Secondly, we take another special value of $s$ which satisfies
$M_{\phi}^2-s=M_{\phi}\Gamma_{\phi}$, we obtain
 \beq
 \left|1-\frac{1+i}{2}Z\right|^2 = \left|1-\frac{1+i}{2}Z'\right|^2~ ,
 \eeq
or
 \beq
 \label{eq5}
 |Z|^2-2\Re(Z)+2\Im(Z)=|Z'|^2-2\Re(Z')+2\Im(Z')~.
 \eeq
Subtraction of Eq.~(\ref{eq3}) from Eq.~(\ref{eq5}) yields
 \beq
 \label{eq6}
 \Re(Z')=\Re(Z)~.
 \eeq
With this equality, Eq.~(\ref{eq3}) is recast as
 \beq
 \label{eq7}
 [1+\Im(Z')]^2=[1+\Im(Z)]^2~,
 \eeq
by virtue of which one gets either $\Im(Z')=\Im(Z)$ or
$\Im(Z')=-2-\Im(Z)$. As a summary, we have two sets of solutions:
 \beq
\begin{array}{rcl}
\Re(Z')&=&\Re(Z)~, \\
\Im(Z')&=&\Im(Z)~;
\end{array}
\label{solution1}
 \eeq
and
 \beq
\begin{array}{rcl}
\Re(Z')&=&\Re(Z)~, \\
\Im(Z')&=&-2-\Im(Z)~.
\end{array}
\label{solution2}
 \eeq
It is readily to check that the above two sets of solutions are
true for the relation~\eref{gsqrelation} for any $s$. Obviously,
the first set of solution is trivial which can be expected
intuitively. However, the second set of solution is fairly
interesting which is firstly obtained analytically. More
interesting thing is, according to the Eqs.~\eref{solution1} and
\eref{solution2}, the second set of solution can be obtained from
the first one. Both solutions describe the experimental data
identically well and one can not distinguish them purely from the
experimental data. Therefore we conclude that if the cross section
of $\EE\to \omega\piz$ as a function of the center-of-mass energy
is parameterized as Eq.~(\ref{xsection}), there must be two sets
of solutions of the interference parameter $Z$.

One remark on our mathematical analysis. More generally we write
$G(s,Z)$ in the form
 \beq
G(s,Z) =1+Z F(s)~,
\label{defggsz}
 \eeq
with $F(s)$ being a complex function depending on $s$. If two
special values of $s$ are taken, from relation~\eref{gsqrelation},
we obtain two quadratic equations with two unknowns. Generally
speaking, there should be four sets of solutions. The existence of
the exact two sets of solutions of our example indicates that the
number of sets of solutions depends strongly on the form of
$F(s)$.

\section{Experimental confirmation}

From Ref.~\cite{kloe_omegapi}, $\Re(Z)=0.106$ and $\Im(Z)=-0.103$
are acquired from a fit to the experimental data in the $\omega\to
\pp\piz$ decay mode. In the light of Eq.~\eref{solution2}, the
second set of solution can acquired immediately, i.e.
$\Re(Z)=0.106$ and $\Im(Z)=-1.897$.

It is interesting to compare these results with those obtained
from a fit to the experimental data~\cite{Yuan:2009gd}. From
Table~2 of Ref.~\cite{Yuan:2009gd}, we found that the real parts
of the two solutions are identical, while the imaginary parts sum
up to a number slightly different from $-2$. A check of the fit
results showed in Ref.~\cite{Yuan:2009gd} indicates that the
$-1.90$ is indeed from a rounding of $-1.897$. Keeping one more
digit, we find the sum of the imaginary parts is exactly $-2$.
Both the real parts and the imaginary parts agree perfectly
between the fit results and the analytical evaluation.

\section{Summary and discussions}

We showed above that there must be two solutions in extracting the
branching fraction of $\phi\to \omega\piz$ with the
parametrization of the $\EE\to \omega\piz$ cross section around
the $\phi$ resonance in Ref.~\cite{kloe_omegapi}. While the first
solution corresponds to $\BR(\phi\to \omega\piz)=4\times 10^{-5}$
as reported in Ref.~\cite{kloe_omegapi}, the second solution would
be $\BR(\phi\to \omega\piz)=7\times 10^{-3}$, which is two orders
of magnitude higher than the first one.

One may need to check whether the parametrization of the cross
section is meaningful or if there are further constraints to the
parametrization or the parameters, in order to pick out the
physics solution from the two-fold ambiguities.

It is worth pointing out that $\phi\to \omega\piz$ is an isospin
violating process and thus should be small. The branching fraction
reported in Ref.~\cite{kloe_omegapi} is already large compared to
theoretical calculations~\cite{zhaoq}. However, if the physics is
the second solution showed above, we would find theoretical
calculations are too low.

\acknowledgments

This work is supported in part by the National Natural Science
Foundation of China (10775412, 10825524, 10935008), the Instrument
Developing Project of the Chinese Academy of Sciences (YZ200713),
Major State Basic Research Development Program (2009CB825203,
2009CB825206), and Knowledge Innovation Project of the Chinese
Academy of Sciences (KJCX2-YW-N29).


\end{document}